\documentclass[preprint,floats,aps,showpacs,epsf]{revtex4}
\usepackage{amssymb}

\usepackage{epsfig}

\newcommand{\be}{\begin{equation}}
\newcommand{\ee}{\end{equation}}
\newcommand{\bea}{\begin{eqnarray}}
\newcommand{\eea}{\end{eqnarray}}

\begin{document}
\title{Low temperature correlation functions in integrable models: Derivation of the large
distance and time asymptotics from the form factor expansion:  see the new version on cond-mat/0508618}

\author{B. L. Altshuler$^*$, R. M. Konik$^\dagger$, and A. M. Tsvelik$^\dagger$}
\affiliation{$^*$ Physics Department, Princeton University, Princeton NJ 08544, 
USA;  Physics Department, Columbia University, New York, NY 10027; NEC-Laboratories America, Inc., 4 Independence Way, 
Princeton, NJ 085540,USA;\\ 
 $^\dagger$Department of  Physics, Brookhaven National Laboratory, Upton, NY 11973-5000, USA}
\date{\today}

\begin{abstract}
We propose an approach to the problem of low but 
finite temperature dynamical correlation functions in integrable 
one-dimensional models with a spectral gap. The approach is based on the analysis of the leading 
singularities of the operator matrix elements and is not model specific. We discuss only models 
with well defined asymptotic states. For such models the long time, large distance asymptotics of 
the correlation functions fall into two universality classes. These classes differ primarily by whether the 
behavior of  the two-particle S matrix at low momenta is
diagonal  or corresponds to pure reflection.  We discuss similarities and differences between our 
results and results obtained by the semi-classical method  suggested by Sachdev and Young, Phys. 
Rev. Lett. {\bf 78}, 2220 (1997).

\end{abstract}
\maketitle
 The earlier version of the paper is replaced with cond-mat/0508618.

\end{document}